\documentclass[twocolumn]{aastex63}
\pdfoutput=1

\usepackage{amsmath}
\renewcommand{\thesubsection}{\Roman{subsection}}

\shorttitle{Proton energization by phase-steepening of parallel propagating Alfv\'enic fluctuations}
\shortauthors{González et al. 2021}
\graphicspath{{./}{figures/}}

\begin{document}

\title{Proton energization by phase-steepening of parallel propagating Alfv\'enic fluctuations}

\correspondingauthor{C.A. González}
\email{carlos.gonzalez1@austin.utexas.edu}

\author{C.A. González}
\affiliation{Department of Physics, The University of Texas at Austin, Austin, TX 78712, USA}

\author{A. Tenerani}
\affiliation{Department of Physics, The University of Texas at Austin, Austin, TX 78712, USA}

\author{L. Matteini}
\affiliation{Department of Physics, Imperial College London, London SW7 2AZ, UK}

\author{P. Hellinger}
\affiliation{Astronomical Institute, CAS, Bocni II/1401, Prague CZ-14100, Czech Republic}
\affiliation{Institute of Atmospheric Physics, CAS, Bocni II/1401, Prague CZ-14100, Czech Republic}

\author{M. Velli}
\affiliation{Department of Earth, Planetary, and Space Sciences, University of California, Los Angeles, CA, USA}

\begin{abstract}
Proton energization at magnetic discontinuities generated by  phase-steepened fronts of parallel propagating, large-amplitude Alfvénic fluctuation is studied using  hybrid simulations. We find that dispersive effects yield to the collapse of the wave via phase steepening and the subsequent generation of compressible fluctuations that mediate an efficient local energy transfer from the wave to the protons. Proton scattering at the steepened edges causes non-adiabatic proton perpendicular heating.  Furthermore, the parallel electric field at the propagating fronts mediates the acceleration of protons along the mean field. A steady-state is achieved where proton distribution function displays a field-aligned beam at the Alfv\'en speed, and compressible fluctuations are largely damped. We discuss the implications of our results in the context of Alfv\'enic solar wind.     
\end{abstract}


\section{Introduction} \label{sec:intro}

In-situ spacecraft measurements show that the solar wind is permeated by nearly incompressible ($\delta |{\bf B}|/{|\delta \bf B}|\ll 1$), large amplitude ($|\delta {\bf B}|/{\bf B}_0\sim 1$) fluctuations \citep{coleman1967} in the velocity and magnetic field which are correlated mainly in the sense of  Alfvén waves propagating away from the sun \citep{belcher1971, Damicis}, and that are characterized by a nearly constant magnetic field magnitude, a condition corresponding to spherical polarization \citep{BRUNO2001,Matteini2015}. Alfv\'enic fluctuations in the solar wind  thus display a high degree of coherence that manifests itself not just by the velocity-magnetic field correlation that characterizes Alfv\'en waves, but also via an intrinsic degree of phase coherence among the oscillating fields that is necessary in order to maintain a locally constant-$B$ constraint. Such Alfv\'enic fluctuations also display typically turbulent features including a well developed energy spectrum and the ubiquitous presence of intermittent structures, that in turn provide suitable places where energy dissipation and particle energization can occur \citep{Marsch_2006,Osman_2010,Tessein_2013,Perrone_2016}.

Rotational discontinuities  are typically found  at the steepened edges of arc-polarized Alfv\'en waves, a particular case of constant-B fluctuations in a 1D  geometry~\citep{Barnes&Hollweg1974,TsurutaniEA2005,erofeev2019characteristics}. It is believed that rotational discontinuities are necessarily generated by wave steepening and the occurrence of abrupt changes in the wave phase \citep{Cohen&Kulsrud2001,MedvedevEA97,malara_1991,Vasquez2001,TsurutaniEA2018,Valentini_2019}. However, how are compressible effects quenched and the nearly constant-$B$ condition maintained during the dynamical phase wave steepening still remains elusive, especially when the plasma beta is smaller than unity and strong couplings with compressible modes are expected~(e.g., \cite{malara_1991,jayanti_1993, roberts_1995,malara_Phys_fluids_1996}). On the other hand, Alfv\'enic fluctuations are sustained by a non-thermal plasma that displays temperature anisotropies and preferential perpendicular heating, and a stable, field-aligned proton beam streaming ahead of the proton core population at the local Alfv\'en speed \citep{marsch1982,matteini2013,Sorriso-ValvoEA2019,Verniero_2020}.

It is the goal of this letter to investigate  the connection between Alfv\'en wave steepening, plasma compressibility, and the development of non-thermal features in a low-beta plasma. It has been shown via hybrid simulations that a large amplitude monochromatic Alfv\'en wave is subject to parametric instabilities \citep{araneda2008proton,MatteiniEA2010,nariyuki2009,gonz_lez_2020}. The decay of the wave leads non-linearly to enhanced proton heating and to a field-aligned beam driven by the field-aligned electric field generated during the wave decay.  Here, we consider the more realistic situation of a broad-band Alfv\'enic fluctuation and adopt a hybrid framework complemented by test-particle simulations to investigate the connection between Alfv\'en wave dynamics and proton energization. In this case, due to the presence of multiple wavelengths and dispersion, the initial fluctuation undergoes phase steepening resulting in the rapid collapse of the wave \citep{spangler1989,ButiEA2001}. We show that phase steepening  and wave collapse is accompanied by the formation of rotational discontinuities embedded in compressional structures at steepened fronts. Proton heating and acceleration occur locally at the steepened fronts, and ultimately contribute to the dissipation of compressible fluctuations.  
\section{Model and Simulation setup}
\label{sec:theory}

In this work we adopt a hybrid model of the plasma that describes electrons as a massless and isothermal fluid and protons as particles via the (non-relativistic, quasi-neutral) Vlasov-Maxwell's equations:

\begin{subequations}
    \begin{equation*}
        \frac{ \partial f_i } {\partial t} + \textbf{v} \cdot \frac{\partial f_i}{\partial \textbf{r}} + \frac{e}{m_i}\left( \textbf{E} +  \frac{\textbf{v}_i}{c} \times \textbf{B} \right) \cdot \frac{\partial f_i}{\partial \textbf{v}} = 0
        \label{eq:New1} 
    \end{equation*}
    \begin{equation*}
        \frac{\partial \textbf{B}}{\partial t} =  - c \  \nabla \times \textbf{E}, \ \ \ \ \ \textbf{J} = \frac{c}{4 \pi} \nabla \times \textbf{B}
        \label{eq:Max1}
    \end{equation*}
    \begin{equation*}
    \textbf{E} + \frac{\textbf{u}_i}{c} \times \textbf{B} =   - \frac{k_B T_e \nabla n }{ e n}  + \frac{ \textbf{J} \times \textbf{B}}{ e n}  + \eta \nabla \times \textbf{B},
    \label{ohm}
\end{equation*}
    \label{Eq1}
\end{subequations}

with $c$ the speed of light, $e$ the electron charge, $k_B$ the Boltzmann constant and $T_e$  the electron temperature. The proton number density $n$ and the proton bulk velocity $\textbf{u}_i$ are computed from the moments of the distribution function ($n = \int{f(\textbf{r},\textbf{v},t) d\textbf{v} }$ and $n \textbf{u}_i = \int{ \textbf{v} f(\textbf{r},\textbf{v},t) d\textbf{v}}$ respectively). Since the model assumes quasi-neutrality, the electric field is obtained through the generalized Ohm's law where contributions from inductive, Hall, electron pressure and resistive terms are retained while electron-inertia and higher-order terms are not considered.

We performed 2.5D simulations by means  of the hybrid-PIC code CAMELIA (e.g., \citet{Franci_2018}). Periodic boundary conditions are imposed in all directions of the computational box. Lengths are normalized to the proton inertial length $d_i =  c/\omega_{p}$ with $\omega_p = (4\pi ne^2/m_i)^{1/2}$ the proton plasma frequency. Time is normalized to the inverse of the proton gyrofrequency $\Omega_{ci}^{-1} = (eB_0/m_i c)^{-1}$ and velocities  to the Alfv\'en speed $v_A = B_0/(4 \pi n m_i)^{1/2}$.  The plasma beta for both ions and electrons is defined as $\beta_{p,e}=8 \pi n k_B T_{p,e}/B_0^2$. We have included explicit resistivity to improve energy conservation by avoiding energy accumulation at the grid scales. The associated dissipation length scale is chosen to be greater than the grid size but smaller than proton scales. 
We start from an initially uniform and isotropic plasma. We initialize the system with an exact nonlinear solution of the MHD system corresponding to a large scale, non-monochromatic and constant-B Alfv\'enic fluctuation propagating along the mean magnetic field ${\bf B}_0$, taken along the \textit{x-axis}. The magnetic field of the wave is  given by  $\delta b_z = \delta b_0 \cos{(\phi(k_0,x))}$ and $\delta b_y = - \delta b_0 \sin{(\phi(k_0,x))}$, with $|\delta {\bf b}|=\delta b_0$ the amplitude of the  wave normalized to the mean magnetic field magnitude $B_0$. The phase $\phi(k_0,x) = k_0 x + \epsilon \sum_{m=n_i, m \neq n_0}^{n_f}  \frac{k_0}{k_m} \cos{(k_m x + \phi_m)}$, where $\phi_m$ is a random phase between $ [0,2 \pi)$. The main wave vector is $k_0 = 2\pi n_0/L$ and the initial wave satisfies the Walen relation in the dispersion-less limit $\delta {\bf u} = -(\omega_0/k_0) \delta {\bf b}$.  The wave frequency $\omega_0$ is determined from the normalized dispersion relation $k_0^2=\omega_0^2/(1-\omega_0)$ for left-handed circularly polarized waves in parallel propagation. This initial condition corresponds to a broad-band Alfv\'enic fluctuation comprised of modes ranging from $k_i=2 \pi n_i/L$ to $k_f=2 \pi n_f/L$. The  parameter $\epsilon$ controls the deviation from the monochromatic case which is recovered for $\epsilon=0$ \citep{malara_Phys_fluids_1996}.  

We present results for a squared box simulation of side  $L=128 d_i$ with 1024 particles per cell and  $1024^2$ grid points, corresponding to a grid size $\Delta  = d_i/8$.  We set the time step to move the particles as $\Delta t=0.01 \Omega_{ci}^{-1}$ and used $30$ substeps for the fields. The main wave vector is $k_0d_i=0.2$ and we set $\epsilon=0.5$. This yields an initial spectrum  with spectral index $-2$ in the interval $k d_i=[0.05,0.5]$. The plasma beta is $\beta_{p,e}=0.5$ and the  wave amplitude $\delta b_0=0.5$. Finally, the explicit resistivity is $\eta=0.0002$. Throughout the text we will use a field-aligned coordinate system and will refer to parallel and perpendicular in terms of the direction of the total magnetic field $\hat{\textbf{b}} = \textbf{B}/\Vert \textbf{B} \Vert$. The temperatures are defined in terms of the decomposition of the pressure tensor according to the direction of the total magnetic field as: $p_\parallel = \textbf{p} \colon \hat{\textbf{b}}\hat{\textbf{b}}$ and  $p_\perp = \textbf{p} \colon (\mathbb{\textbf{I}} - \hat{\textbf{b}}\hat{\textbf{b}})/2$ and the pressure tensor $\textbf{p} = \int(u-v_p)_i(u-v_p)_j f({\bf r,v},t) d{\bf v}$ is obtained from the particle velocity distribution. 

\section{Results} \label{sec:results}

\begin{figure}
\includegraphics[width=0.5\textwidth]{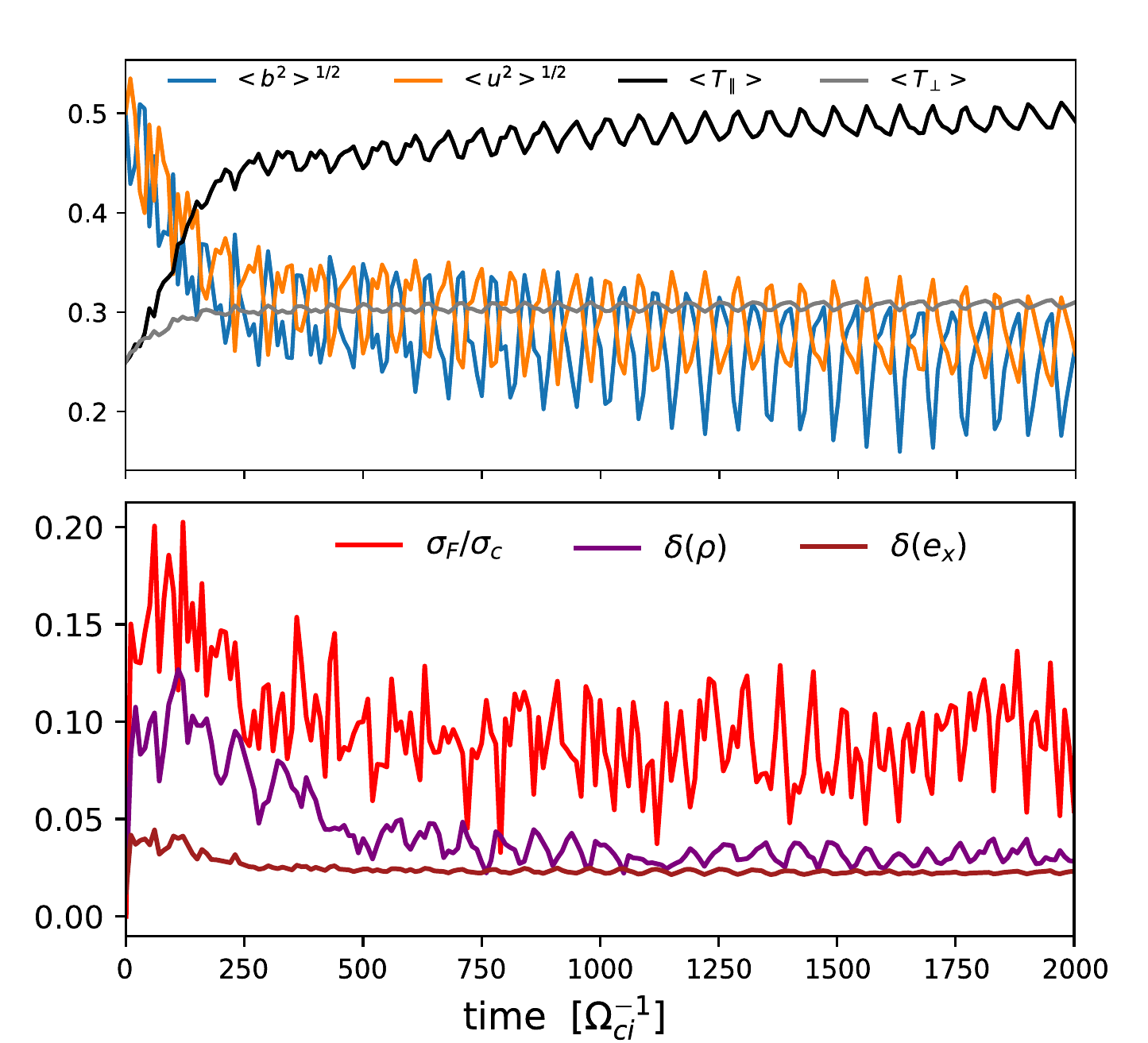}
\caption{(\textbf{Top}). Temporal evolution of the rms magnetic and velocity fluctuations and  mean proton temperature decomposition. (\textbf{Bottom}). Time evolution of the ratio $\sigma_F/\sigma_c$ and the standard deviation of the density and the field-aligned electric field.}
\label{Fig1}
\end{figure}

Fig.~\ref{Fig1} summarizes the time evolution of the system and the top panel shows the root mean square (rms) of the magnetic and velocity field fluctuations, and also the average parallel and perpendicular proton temperatures. The initial Alfv\'enic fluctuation collapses after a few proton gyroperiods by releasing its energy to the plasma in the forms of thermal and kinetic energy, the latter via the formation of a field-aligned beam, as we discuss below, until a steady-state is achieved at around $250 \ \Omega_{ci}^{-1}$.

The disruption of the wave and the resulting proton energization and beam formation are due to the phase-steepening of the initial wave. Because of dispersive effects, we observe a rapid phase steepening of the wave due to the larger scales catching up with the smaller ones, ultimately leading to a modulation of the magnitude of $|{\bf B}|$ and hence to localized steepened wave fronts, a process analogous to the collapse of localized Alfv\'enic wave packets due to modulational instability~\citep{ButiEA2001, spangler}. Departures from the initial constant-B state drives compressible fluctuations and a field-aligned electric field at the steepened edges of the Alfv\'enic fluctuation.

The compressive fluctuations are displayed in the bottom panel of Fig.~\ref{Fig1}, where the standard deviation ($\delta (X) = \ < \left(  X  - \left<  X \right> \right)^2 >^{1/2}$) of proton density and the field-aligned electric field are plotted.  The ratio of the standard deviation of the magnetic field magnitude ($\sigma_F = \delta(|{\bf B}|)$) to the magnetic field fluctuations ($\sigma_c = [ \delta^2(b_x) + \delta^2(b_y) + \delta^2(b_z)]^{1/2}$) is also shown. All These quantities display a maximum during the initial stage of wave collapse/disruption and then minimize at saturation, pointing to the fact that nonlinear wave-particle interactions at the steepened edges reduce plasma compressibility.  

\begin{figure}
\includegraphics[width=0.45\textwidth]{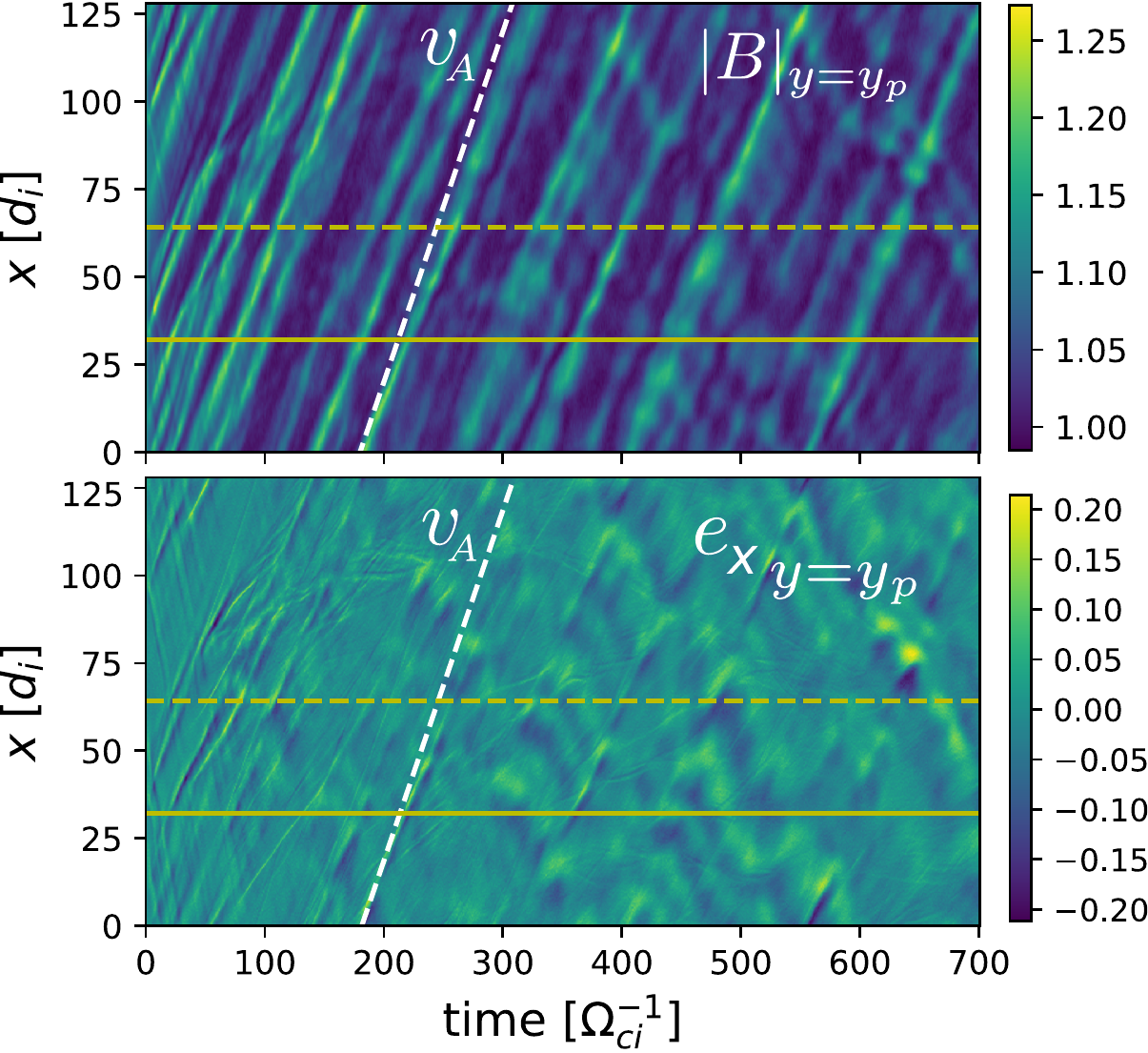}
\caption{Contour plot in the $(x-t)$ plane of the magnitude of the magnetic field (\textbf{Top}) and of the parallel electric field  (\textbf{Bottom}) at $y_p=64 d_i$. The yellow lines mark the location of the two points used to reproduce Fig~\ref{Fig4}. As a reference, we show in white dashed lines the characteristic propagating at the Alfvén velocity.}
\label{Fig2}
\end{figure}

In Fig.\ref{Fig2} we show a contour plot (spatio-temporal diagram \textit{x-t} plane ) of the magnitude of the magnetic field and the mean field-aligned electric field at the plane $y_p=64 \ d_i$, showing the characteristics of the discontinuities generated at the wave edges as the result of phase-steepening. These structures propagate at nearly the Alfv\'en speed and essentially along the mean magnetic field. One can observe that the shape of these structures changes as they propagate, and that they largely fade away at saturation (at around $t=300 \  \Omega_{ci}^{-1}$).  The main contribution to the field-aligned  electric field at those locations comes from the Hall term ($\textbf{J} \times \textbf{B} = \textbf{B} \cdot \nabla \textbf{B} - \nabla(\textbf{B}^2/2)$), in particular from the field-aligned component of the second term on the right hand side \citep{MatteiniEA2010,gonz_lez_2020}. The interaction of protons with these propagating discontinuities and the parallel electric field therein provides a suitable mechanism for the acceleration and heating of protons. 

\begin{figure}
\includegraphics[width=0.45\textwidth]{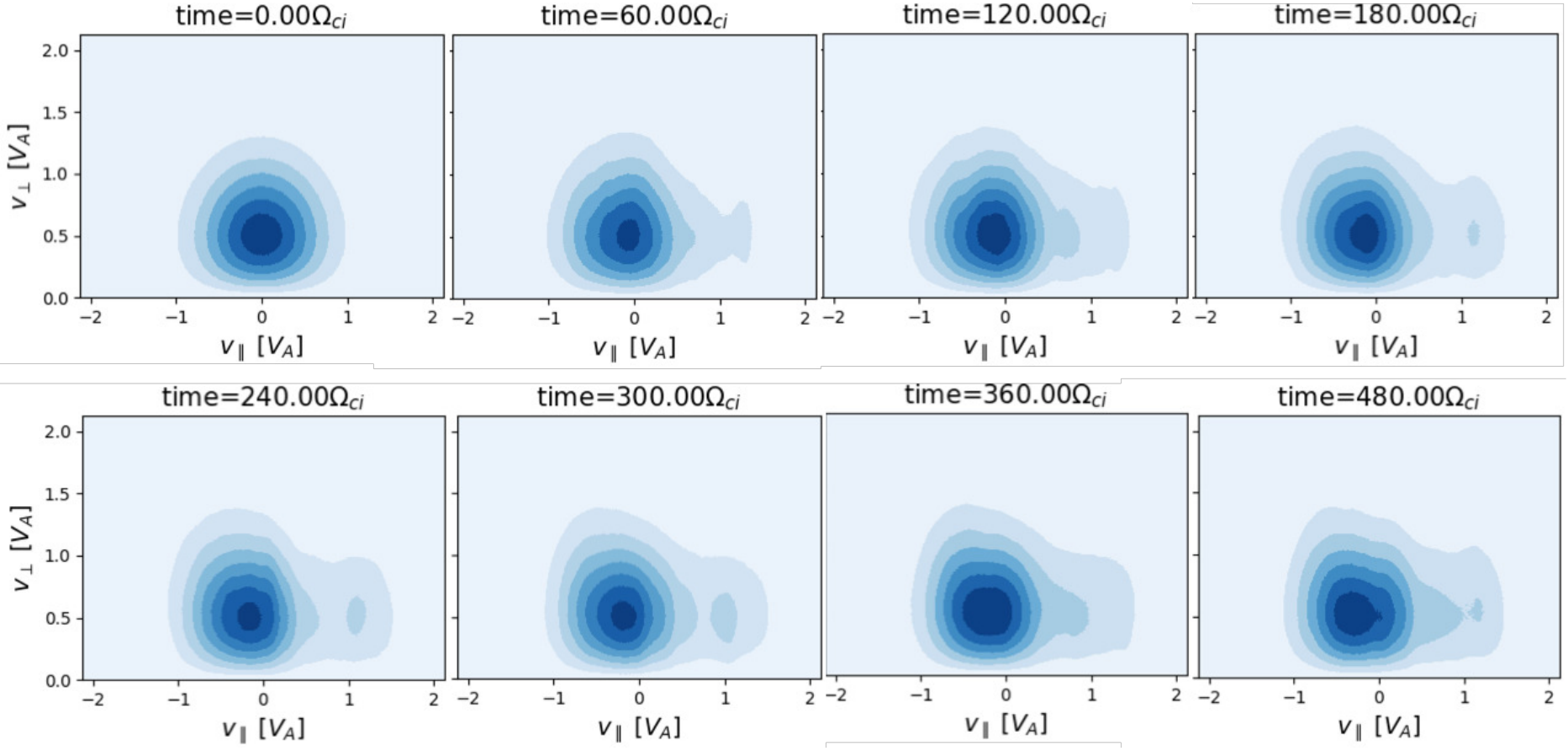}
\caption{Contours of the proton velocity distribution function in the $(v_\parallel-v_\bot)$ plane at different times.}
\label{Fig3}
\end{figure}

The proton velocity distribution function (VDF) in the $(v_\parallel-v_\bot)$ plane is shown in Fig~\ref{Fig3}. The VDF is displayed early in the evolution when the strongest heating  is taking place. As can be seen, the largest contribution to the  parallel temperature comes from the field aligned proton beam at the Alfvén speed. The proton beam is generated early in the evolution after the formation of magnetic pressure fluctuations and remains stable throughout the final steady-state. A qualitative similar behavior was found in \cite{nariyuki_2014} by means of a reduced one dimensional MHD-Vlasov model in a radially expanding geometry. In their simulations a beam also forms at steepened wave fronts generated by MHD nonlinearities, rather than by dispersive effects.

\begin{figure}
\includegraphics[width=0.45\textwidth]{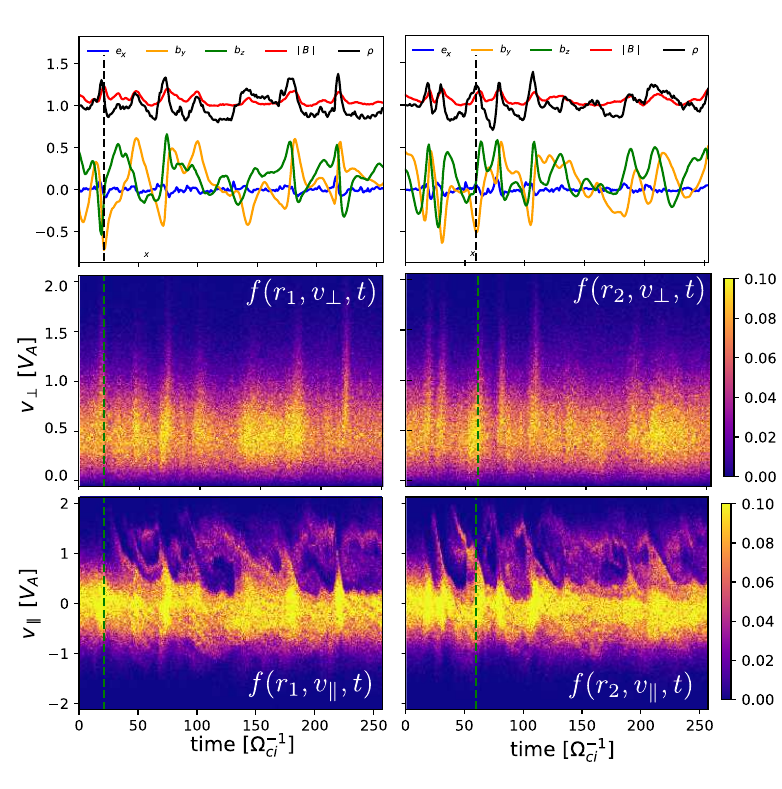}
\caption{Single point measurements at position $r_1$ (\textbf{Left}) and $r_2$ (\textbf{Right}). The top panels show the time series of $b_y$, $b_z$, the parallel electric field $e_\parallel$, the magnitude of B and the particle density $\rho$. The middle and bottom panels show the time series of the reduced proton VDF in the perpendicular and parallel directions, respectively. }
\label{Fig4}
\end{figure}

To illustrate the dynamical phase-steepening process and its implications on the particles VDF, we computed single-point measurements of some field quantities and the reduced parallel and perpendicular VDF at two different fixed points in the simulation domain. The two probes are located at  $r_1 = (32 \ d_i,64 \ d_i)$ and $r_2 = (64 \ d_i,64 \ d_i)$, respectively, and correspond to the yellow lines marked in Fig~\ref{Fig2}. The time series at both probes is presented in Fig~\ref{Fig4}, where we show the transverse magnetic field fluctuations, the magnitude of the magnetic field, the particle density, and the field-aligned electric field as a function of time. 
The vertical dashed lines in the top panels of Fig~\ref{Fig4} show the location of a single structure that crosses the two points. We estimated that the structure propagates at a speed  $V=0.82 \ v_A$, although the characteristics in Fig.~\ref{Fig2} show that the speed is not constant and some of the steepened fronts break down. 

The resulting signatures on the proton VDF due to rotational discontinuities on the particle VDF are different for the parallel and perpendicular components. Plume-like structures can be identified in the perpendicular component with enhancement of particles with large $v_\bot$ at the location at the steepened fronts. On the other hand, the velocity-space structures in the parallel component show the presence of beams in front of the discontinuities. As the faster particles in the beam are farther from the steepened fronts, they arrive at the probe earlier, hence the tilted structure that can be seen in the time series of the parallel VDF. Besides, phase-space holes and slower parallel particle velocity right in the location of the steepened fronts are evidenced.

To understand the energization process produced by the interaction of protons with the rotational discontinuities, we computed the trajectory of test-particles using high-cadence electric and magnetic field data generated by the simulation. The initial test-particle ensemble is randomly distributed through the simulation box with initial Maxwellian distribution function for perpendicular velocities while particles are initialized with zero velocity along the mean field. Periodic boundary conditions are imposed in \textit{x-y} directions and we considered protons with the same physical parameters as in the hybrid simulation.

\begin{figure}
\includegraphics[width=0.45\textwidth]{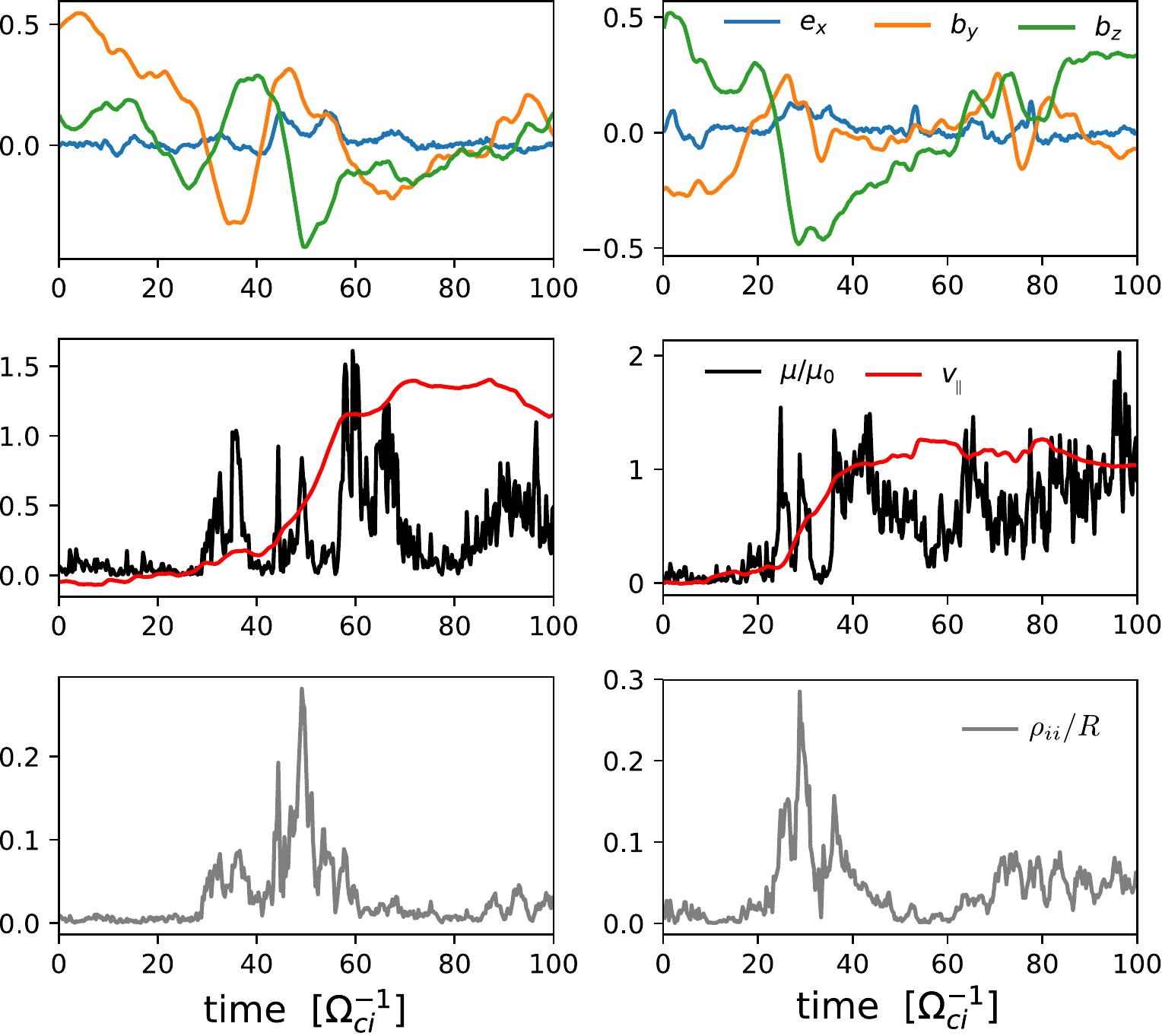}
\caption{Particle information along the path of two representative test-particles. (\textbf{Top}). The electric ($e_x$)and magnetic field components ($b_y$ and $b_z$). (\textbf{Middle}). The parallel  particle velocity $v_\parallel$ and the normalized particle magnetic moment $\mu/\mu_0$. (\textbf{Bottom}). It is  shown the ratio between the particle gyro-radii to the radius of curvature of the magnetic field $\rho_{i}/R$.}
\label{Fig5}
\end{figure}

Fig.~\ref{Fig5} shows the tracking of two representative particles that undergo the acceleration process resulting from the interaction with discontinuities. The top panel shows the field interpolated along the particle trajectory. The middle panel shows the magnetic moment $\mu = m_i v_\bot^2/2B_0$, normalized to the  magnetic moment evaluated with the initial proton thermal velocity (black line) and the parallel velocity (red line). Proton interaction with rotational discontinuities results in the violation of the first adiabatic invariant. The non-adiabatic particle behaviour is observed once the particle crosses a discontinuity, followed by the acceleration of the particle along the magnetic field due to the field-aligned electric field inside the structure. The bottom panel of Fig.~\ref{Fig5} presents the ratio between the particle gyroradius $\rho_{i}=v_{th}/\Omega_{ci}$ and the radius of curvature of the magnetic field at particle location $R = 1/\kappa$ , with the curvature defined as  $\kappa =  \Vert \hat{\textbf{b}} \cdot \nabla \hat{\textbf{b}}\Vert$. The breaking of adiabatic particle motion at discontinuities occurs when a particle experiences a sharply curved magnetic field rotation at the discontinuities, which allows the stochastic behaviour of protons in those regions. This is consistent with the signatures on the perpendicular  VDF (middle panel of Fig.~\ref{Fig4}), showing an enhancement of particles with larger velocities each time a magnetic structure is crossed. In conclusion, the proton heating and acceleration process resulting from the phase-steepening of large-amplitude Alfvénic fluctuations is complex. The natural development of parallel propagating structures that travel at around the Alfvén speed  involve a  bipolar electric field that accelerates particles into a mean field-aligned beam, while scattering by the magnetic field structure contribute at the same time to an enhancement of perpendicular heating. Particle that are being accelerated into the beam may resonate with the propagating structure leading to the damping of compressive fluctuations and allowing the final non-linear steady-state.

\section{Conclusions} \label{sec:conclusion}

In this letter we made use of hybrid simulations, complemented by test particle simulations, to investigate proton energization at the phase-steepened edges of an initial large-amplitude, constant-B  Alfvénic fluctuation in a low beta plasma.   We find that dispersion leads to the initial phase steepening of the wave resulting in its rapid collapse, as predicted for weakly dispersive Alfv\'enic wave packets. This process is accompanied by the formation of rotational discontinuities embedded in compressional structures characterized by an enhancement of magnetic pressure at the steepened fronts that propagate at a speed slightly less than the Alfv\'en speed. Proton perpendicular heating via pitch angle scattering and parallel acceleration take place in those localized regions, due to the  interaction of protons with the parallel electric field (mainly induced by the gradients of $|{\bf B}|$) therein. Within the fully self-consistent hybrid simulations, it is those demagnetized protons, accelerated up to about the Alfv\'en speed,  to ultimately mediate the damping of the parallel electric field and reduce compressible fluctuations via nonlinear wave-particle resonance.

Our results provide a possible explanation for the ubiquitous presence of a stable, field aligned proton beam, commonly observed in the Alfv\'enic wind, and enhanced proton heating. We also argue that the existence of the proton beam itself is intimately related to the quenching of compressible fluctuations and to the nearly constant-B field that characterizes the Alfv\'enic wind. Our simulation display a final degree of compressibility of fluctuations comparable to solar wind observations \citep{villante1982}, although some steepened fronts survive or tend to reform. However, further investigations on the role of electrons physics, not included within the hybrid model, is needed, since electrons may contribute significantly to the damping of compressible fluctuations via  resonant interactions \citep{hollweg_1971}.

\acknowledgements{We would like to thank A. Artemyev for useful discussions. This research was supported by NASA grant \#80NSS\-C18K1211 and the HERMES NASA DRIVE Science Center Grant n. 80NSSC20K0604. We acknowledge the Texas Advanced Computing Center (TACC) at The University of Texas at Austin for providing HPC resources that have contributed to the research results reported within this paper. URL: http://www.tacc.utexas.edu.}

\bibliography{sample63}{}
\bibliographystyle{aasjournal}
\end{document}